\documentstyle[preprint,prl,aps,epsf]{revtex}

\begin{document}


{\tighten
\preprint{{\vbox{ \hbox {July 1997} \hbox {rev. Oct. 1997} \hbox{IFP-742-UNC}\hbox{VPI-IPPAP-97-5}}}}
 


\title{On Radiative CP Violation in (Supersymmetric) 
Two Higgs Doublet Models}

\author{\bf Otto C. W. Kong$^a$\footnote{Present address: 
Department of Physics and Astronomy,
University of Rochester, Rochester NY 14627-0171. E-mail: kong@pas.rochester.edu}
 and  Feng-Li Lin$^b$\footnote{E-mail: linfl@vt.edu}}
\address{
$^a$ Institute of Field Physics, Department of Physics and Astronomy,\\
University of North Carolina, Chapel Hill, NC  27599-3255.\\
$^b$ Department of Physics, and Institute for Particle Physics and Astrophysics,\\ 
Virginia Polytechnic Institute and State University, Blacksburg, VA 24061-0435.}

\maketitle

\begin{abstract}
We discuss the feasibility of spontaneous CP violation being
induced by radiative corrections in 2HDM's,
emphasizing on a consistent treatment at 1-loop level. Specifically,
we analyze the cases of gaugino/higgsino effect on MSSM, and a new model
proposed here with an extra, exotic, quark doublet. The latter model is 
phenomenologically interesting. One conclusion is that some fine
tuning is in general needed for the scenario to work. The case
for the new model requires a tree level mass mixing between the
two Higgses, which fits in the standard SUSY picture. The fine tuning
requires is then very moderate
in magnitude and in a way natural.
\end{abstract}
}

\newpage 

{\it Introduction.} The source of CP violation is one of the most important
unsolved puzzles in particle physics. The possibility of CP being a 
spontaneously broken symmetry keeps generating new interest.
The most simple setting for achieving the scenario is in a two-Higgs-doublet
model (2HDM), originally analyzed by Lee\cite{lee}. However, in order to 
avoid flavor-changing-neutral-currents that could result, extra structure
like natural flavor conservation (NFC)\cite{nfc} has to be imposed on
a 2HDM, which then forbids  spontaneous CP violation (SCPV)\cite{brc}.
A supersymmetric version of the standard model (SM) is naturally a 2HDM
with NFC being imposed automatically by the holomorphy 
of the superpotential. Hence, the bulk of study on the   
minimal supersymmetric standard model (MSSM) concentrates
on the CP conserving vacuum. Recently, Maekawa\cite{mae} illustrated
that SCPV could actually arise in the model in some region of the 
parameter space through radiative correction. In a subsequent paper,
Pomarol\cite{pol} argued that  mass for the "psuedoscalar", $m_A$,
is roughly proportional to $(\lambda_5)^{1/2}$ and is at least
more than a factor of three too small to be phenomenologically
acceptable. Both analyses, however, are in a way oversimplified.
For instance, the $m_A$ mass estimate is, strictly speaking, obtained
from a Higgs mass matrix of negative determinant, hence indicating
the whole picture is liable to be totally inconsistent.
In this letter, we  look into the situation more carefully and try to address
the various issues involved in a radiative SCPV model of the kind,
with or without supersymmetry (SUSY). 
The basic idea here is that in order to
have a {\it consistent approximation} to the radiative corrections in
whatever interesting region of the parameter space, loop contributions
to {\it all} parameters in the scalar potential at the same order have 
to be considered. Our result illustrates that,  the radiative CP 
violation scenario {\it does not} get around the Georgi-Pais
theorem\cite{gp} in the way claimed in the literature. Specifically, fine 
tuning {\it is} in general needed for it to work.

Despite the Georgi-Pais theorem and the $m_A$ mass limit constraint
the radiative CP violation scenario still generate interest\footnote{See
Refs.\cite{ck,Ol}, for example.}. In our opinion, a realistic model
implementing the mechanism, hence realizing CP violation as a
pure quantum mechanical effect, is of general interest. It provides
a natural way of explaining the smallness of CP violation and a potentially
different phenomenology of the latter. Within the domain of a
supersymmetric 2HDM specifically, it also provides the only 
possible scenario of SCPV. Such a model is constructed and presented
here.
Our new model, has an extra,  exotic, quark doublets.
It could be phenomenologically-viable, in the sense
that $\lambda_5$ resulted could easily be 
more than an order of magnitude larger,
hence circumventing the  likely small $m_A$ obstacle,
though the full phenomenological features of the model still have 
to be worked out. 
Unlike a recent model  using right-handed neutrinos\cite{ck},
our model achieves a large  $\lambda_5$ without a fine tuning 
in the fermion  (quark) masses,
and is fully compatible with SUSY.
However, the same type of fine tuning, though quite moderate
numerically, as in the MSSM case
is needed to really have radiative CP violation. We expect this feature
to be generic.

{\it The scalar potential and SCPV.}   
The most general two Higgs doublet 
potential is given by
\begin{eqnarray}
V(\phi_1, \phi_2) &=& m_1^2 \phi_1^{\dag} \phi_1 + m_2^2 \phi_2^{\dag} \phi_2 - (m_3^2 \phi_1^{\dag} \phi_2 + h.c.) 
\nonumber\\
&&+ \lambda_1 (\phi_1^{\dag} \phi_1)^2 + \lambda_2 (\phi_2^{\dag} \phi_2)^2 + \lambda_3 (\phi_1^{\dag} \phi_1)(\phi_2^{\dag} \phi_2)
\nonumber\\
&&+ \lambda_4 (\phi_1^{\dag} \phi_2)(\phi_2^{\dag} \phi_1) + \frac{1}{2} \lbrack \lambda_5 (\phi_1^{\dag} \phi_2)^2 + h.c. \rbrack
\nonumber\\
&&+ \frac{1}{2} \lbrace \phi_1^{\dag} \phi_2 \lbrack \lambda_6 (\phi_1^{\dag} \phi_1) + \lambda_7 (\phi_2^{\dag} \phi_2) \rbrack + h.c. \rbrace~.
\end{eqnarray}
Assuming  all the parameters in $V$ being real, and denoting the
vacuum expectation values (VEV's) of the neutral components of the Higgs doublets   by
\[ 
\langle \phi_1^0 \rangle = v_1 \qquad {\rm and}  \qquad
\langle \phi_2^0 \rangle = v_2 e^{i\delta}~,
\]
we have 
\begin{eqnarray}
\left\langle V\right\rangle &=& m_1^2 v_1^2 + m_2^2 v_2^2 + \lambda_1 v_1^4 + \lambda_2 v_2^4 + (\lambda_3 + \lambda_4 - \lambda_5) v_1^2 v_2^2 
\nonumber\\
&&+ 2 \lambda_5 v_1^2 v_2^2 cos^2\delta -(2 m_3^2 - \lambda_6 v_1^2 -\lambda_7 v_2^2) v_1 v_2 cos\delta~,
\nonumber\\
&=& M_1 v_1^2 + M_2 v_2^2 + (p v_1^4 + 2r v_1^2 v_2^2 +q v_2^4)
\nonumber\\
&&+ 2 \lambda_5 v_1^2 v_2^2 (cos\delta -\Omega)-\frac{m_3^4}{2 \lambda_5}~;
\end{eqnarray}
where
\begin{equation}
\Omega = \frac{2m_3^2 - \lambda_6 v_1^2 - \lambda_7 v_2^2}{4 \lambda_5 v_1 v_2}~,
\end{equation}
and
\begin{eqnarray}
M_1 &=& m_1^2 + \frac{\lambda_6 m_3^2}{2 \lambda_5}~,
\\
M_2 &=& m_2^2 + \frac{\lambda_7 m_3^2}{2 \lambda_5}~,
\\
p &=& \lambda_1 - \frac{\lambda_6^2}{8 \lambda_5}~,
\\
q &=& \lambda_2 - \frac{\lambda_7^2}{8 \lambda_5}~,
\\
r &=& \frac{1}{2}(\lambda_3 + \lambda_4 - \lambda_5 - \frac{\lambda_6 \lambda_7}{4 \lambda_5})~.
\end{eqnarray}
A nontrivial phase ($\delta$) then indicates SCPV. 

Let us look at the $\delta$-dependence of $\left\langle V\right\rangle$. 
The extremal condition gives
\begin{equation}
 -4 \lambda_5 v_1^2 v_2^2(cos\delta - \Omega)sin\delta =0 \; ,
\end{equation}
and the stability condition requires
\begin{equation}
\frac{\partial^2 V}{\partial \delta^2}=4\lambda_5 v_1^2 v_2^2 \lbrack cos\delta(\Omega-cos\delta ) - sin^2\delta \rbrack > 0 \; .
\end{equation}
$cos\delta = \Omega$ gives a SCPV solution, provided that 
$\lambda_5 > 0$ and $|\Omega|< 1$.
Actually,  Eq.(2) shows that this is the absolute minimum.
Then one can easily obtain the result that
\begin{eqnarray}
v_1^2=\frac{1}{2}\frac{rM_2 - qM_1}{pq-r^2}~,
\\
v_2^2=\frac{1}{2}\frac{rM_1 - pM_2}{pq-r^2}~.
\end{eqnarray} 
In order for V to have a lower bound, we have the extra constraints
\begin{equation}
p > 0~,\;\;\;\;\;\;\; q > 0~,
\end{equation}
and
\begin{equation}
pq > r^2~,
\end{equation}
which demand that
\begin{equation}
rM_2 > qM_1~, \;\;\;\;\;\; rM_1 > pM_2~.
\end{equation}
Note that if $p>0$, $q>0$ and $r<0$, then $M_1 M_2 <0$ is required
for a consistent solution. 

On the other hand, for  a CP conserving vacuum, $sin\delta = 0$
and the stability condition reads 
\begin{equation}
\pm \lambda_5 (\Omega \mp 1) > 0~,
\end{equation}
with the two signs correspond  to the cases $\delta = 0$ and 
$\pi$ respectively. Only one of them would give a minimum in
whatever region of the parameter space. For instance, $\delta = \pi$,
corresponding to a negative $\tan\beta(=v_2/v_1)$,
is the only minimum for  $\lambda_5>0,  \Omega<-1$ and
$\lambda_5<0,  \Omega>-1$.

Without SUSY, the natural way to impose NFC is to require that only 
one of the Higgs, say $\phi_1$, transforms nontrivially under an extra
symmetry. This means that $m_3^2$, $\lambda_6$ and $\lambda_7$,
and may be  $\lambda_5$ too, all have to vanish. The same result
is obtained from the superpotential of the MSSM, 
except that the soft SUSY breaking
$B$-term gives rise to a nonvanishing  $m_3^2$. All these are about the
tree level scalar potential. The interesting point of concern here is
whether radiative corrections can modify  the  picture.

{\it Radiative CP violation.} For the MSSM case, 
Maekawa\cite{mae} pointed out that there
is a positive contributions to $\lambda_5$ from a finite 1-loop diagram 
(Fig. 1a) involving the gauginos and higgsinos, which could lead to 
SCPV. The small value of $\Delta\lambda_5(\sim g^4/16\pi^2)$ resulting,
however, could lead to phenomenological problem\cite{pol}. The lesson 
here may be that extra, probably vector-like, fermions with appropriate
mixed couplings to both Higgs doublets could make good 
candidate models for the radiative CP violation scenario. The fermions
may be gauginos and higgsinos, or right-handed neutrinos\cite{ck},
or, as illustrated in our new model below, quarks. These fermions
do lead to flavor changing neutral currents which, however, could 
be made to be sufficiently suppressed.

Our new model has an extra pair of vectorlike quark doublets, 
$Q$ and $\bar{Q}$,
with the following couplings
\begin{equation}
{\cal L}_Q = M_Q  \bar{Q} Q + \lambda_Q \bar{t} \phi_1^{\dag} Q\; ,
\end{equation}
as an addition to the SM with two Higgs doublets or MSSM. Note that 
$\phi_1^{\dag}$ is actually $H_d$, the Higgs (super)multiplet that gives
masses to the down-type quarks; and $\phi_2$ is  $H_u$. 
So, $T_3=-1/2$ component of $Q$, denoted by $T$,
has the same charge as the top quark and mixes with it after electroweak
(EW) symmetry breaking. The other part of the doublet is a
quark of electric $5/3$.
The 1-loop diagram, now with the gaugino/higgsino propagators
replaced by that of the quarks, leads to 
$\Delta\lambda_5(\sim 3\lambda_Q^2\lambda_t^2/16\pi^2)$ 
and could be very substantial for large Yukawa couplings. 
Note that $\phi_1$ and $\phi_2$ vertices now have 
$\lambda_Q$ and $\lambda_t$ couplings respectively. If $M_Q$ around
the same order as the EW scale, this can easily get around
the above mentioned phenomenological objection without 
fine tuning of the fermion masses, unlike the case with right-handed
neutrinos. Modification to top quark
phenomenology would then be very interesting. Another interesting
point is that the scalar partner of $\bar{Q}$ in the SUSY case has
exactly the quantum numbers needed to be a leptoquark that can
produce the high-$Q^2$ anomaly at HERA\cite{hera}.
$Q$-$\bar{Q}$ could naturally arise,
for example, as the only extra quarks from some interesting 
models with a SM-like chiral fermion spectra  embedding the three SM 
families in a intriguing way\cite{kg}.

{\it A consistent treatment.} Nevertheless, a $\Delta\lambda_5$
is a necessary but {\it not sufficient} condition for radiative CP 
violation. This is clearly illustrated in our discussion of the scalar 
potential above. For instance, in the MSSM case, if a 
$\Delta\lambda_5$ is taken as the only modification to the tree level
parameters in Eq.(2), the constraint given by Eq.(15) is upset
and the potential has no lower bound  along $cos\delta =\Omega$.
An explicit calculation of the $3\times 3$ physical Higgs mass squared
matrix $m^2_{Hij}$  actually gives
\begin{equation}
det(m_H^2)=\lambda_5 (pq-r^2)sin^2 2\beta sin^2\delta \; ,
\end{equation}
which  is negative. This inconsistency is also pointed out by Haba\cite{hb}, who then suggested that it would be fixed when top/stop
loop contributions to the other parameters,
mainly $\lambda_2$, in the potential are included.
In our opinion, it is at least of theoretical interest to see  
what the 1-loop gaugino/higgsino effect alone could do to the vacuum
solution of a supersymmetric 2HDM. A consistent treatment of
the 1-loop effect should of course take into consideration 
contributions to all the 10 parameters in the potential $V$.
In Fig.1, we illustrate all the corresponding 1-loop diagrams
involved. Obviously, $\lambda_5$, $\lambda_6$, $\lambda_7$ 
as well as (the sum of)
$\lambda_3$ and  $\lambda_4$ get finite contributions while
the other five parameters receive divergent contributions. 
Note that in the diagrams, we do not distinguish 
$\Delta\lambda_3$ and  $\Delta\lambda_4$; only
their sum is involved in $\left\langle V\right\rangle$. Here we are not going to list
all the tedious analytical expressions of these results, 
we plot the numerical results of major interest in Fig.2 instead.
The plots are for the chargino contributions only, as functions of
$m=M_{\tilde{g}}/\mu$, the gaugino-higgsino mass ratio.
Our results here presented give the 1-loop effect before EW
symmetry breaking, {\it i.e.} mass mixing between the gaugino and 
higgsino were not considered.  Further modifications due to the
symmetry breaking are not expected to change the general features. Here, the
neutralino part can simply be inferred from symmetry. For each charged   
fermion loop diagram, there is one from the neutral $SU(2)_L$
partner of identical amplitude. Then there is another one with
the   $SU(2)_L$  gaugino and the gauge coupling replaced 
by the $U(1)_Y$ ones.   

Now we are in position to discuss the feasibility of radiative CP 
violation within a consistent approximation. In the model,
$\Delta\lambda_6$ and $\Delta\lambda_7$ are  identical. 
For most region of plot,  they are of magnitude
substantially larger than $\Delta\lambda_5$. A essential condition for
SCPV is that $\Omega$, as given by Eq.(3), has magnitude 
less than one. Here the value of a divergent $\Delta m_3^2$ is also
involved. If we take the finite result from $\overline{MS}$ scheme,
for example, $\Delta m_3^2$ has a magnitude that increases fast with
that of $m$ for $|m|>1$.  Obviously, some fine tuning is needed to get
$|\Omega|<1$, though a small window on $m$ always exists not too far
from $|m|=1$, for each sign of $\mu$, for a not too small  $\mu$. 
With EW-scale  $\mu$, 
$\Delta m/m$ of the admissible regions are of order $10^{-2}$, though
the severe fine tuning can be tamed by having small $\mu$. Apart from
the small "pseudoscalar" mass objection, the admissible $\Omega$
region obviously gives phenomenologically unattractive  $\mu$
and $M_{\tilde{g}}$ values. 
Renormalization for $m_3^2$ here could be tricky, as there is
no definite guideline on what its renormalized value should be,
especially in the CP violating case. In a softly broken SUSY model, 
such as MSSM, tree level value for $m_3^2$ is actually
allowed, but then the $B$-term involved gives a negative
contribution to  $\Delta \lambda_5$. Apparently, one can take the
renormalized value of  $m_3^2$ as  a free parameter and adjust it to give
$|\Omega|<1$ for a fix $m$, at least when the $B$-term 1-loop effects
are neglected.  In any case, however, the fine tuning
conclusion stands, as suggested by  the Georgi-Pais theorem. 

The case of our new model is relatively complicated. First of all, the
gaugino/higgsino contributions are there before EW symmetry breaking,
as presented above. In our new model, though the new quark doublet
$Q$ has mass before EW symmetry breaking, a similar set of
1-loop diagrams, as those given in Fig.1, can only be completed when the
EW breaking mass of the top and its mixing with $T$ are taken into 
consideration. But then the plausibly large Yukawa couplings
give rise to substantial results. Otherwise, the qualitative features
are similar to the gaugino/higgsino case, with the exception that
here $\lambda_3 + \lambda_4$ also receives logarithmic 
divergent contributions. In Fig.3, we presented some numerical
results. The plots use again $m$ as parameter which in this case
denotes $M_Q/v_2$. For simplicity, we assume $\lambda_Q = \lambda_t$.
Note that around the CP violating vacuum
of concern here,  $\tan{\beta}\sim 1$ is to be expected in the SUSY
case, as one can easily see by using Eqs.(11) and (12), with tree-level
values of the various parameters assumed.

Here all the results are independent of the sign of $m$.  
$\Delta\lambda_6$ and  $\Delta\lambda_7$ are different but always of
the same signs. They are still of substantially larger magnitude when
compared with $\Delta\lambda_5$, while the latter is of interesting
value for small value of $m$, say $<1$. However, $\Delta m_3^2$
($\overline{MS}$) always has the opposite sign to that of
$\Delta\lambda_6$ or $\Delta\lambda_7$, making a naive use of the
value to fit in the $|\Omega|<1$ condition impossible.
Hence, for SCPV to occur, a tree level  $m_3^2$ value is needed. 
The general feature is independent of $\tan{\beta}$; but making 
its value really large suppresses  $\Delta\lambda_5$ and is hence
undesirable. Soft SUSY breaking also
provides the natural source of a tree level  $m_3^2$, while the $B$-term
loop effects would be relatively negligible as they are suppressed by the small
gauge couplings.  While the  $m^2_{3(tree)}$ then has to
be chosen to roughly match the $t$-$T$ 1-loop effect, the large
Yukawa couplings make this relatively natural, as a value of the order 
$v_2$ is all required. Accepting that,   $\Delta m^2_{3(tree)}/m^2_{3(tree)}$ 
of the solution region is not quite small, 
$\mathrel{\raise.3ex\hbox{$>$\kern-.75em\lower1ex\hbox{$\sim$}}} .35$ 
for $m\leq 1$, representing only a moderate fine tuning. 
The small $m_3^2$ required is natural, as its zero limit provides
the scalar potantial with an extra Peccei-Quinn type symmetry.
We expect a SUSY version of our model
to be phenomenologically interesting. It has extra quarks close to the
top mass scale and can give a feasible radiative CP violation scenario,
though compatibility with top-phenomenology awaits detailed 
investigation.

{\it Further discussions.} An alternative, and may be more
powerful, approach to the problem is to analyse the full 1-loop
effective potential. Actually, we did check explicitly that we get
the same numerical results for the modifications in all the potential
parameters, from differentiating the
standard Coleman-Weinberg expression of the 1-loop potentials
for both cases above. With proper treatment of renormalizations
for all the divergent quantities taken, the approach is
expected to give a much better approximation to the quantum
behavior of the potential and the physical Higgs masses. However, a
detailed Coleman-Weinberg analysis of a 2HDM after symmetry
breaking needed to be done more carefully, for instance, along 
the line suggested in Ref.\cite{cw}. 

Renormalization group runnings and contributions from other
ingredients such as scalars should also be taken into consideration
in the analysis of the detailed phenomenological implications and
feasibility of a full model. The latter is especially important in
the SUSY case. Recall that cancellations between superpartners
is the natural mechanism that eliminates the quadratic divergence,
as those arise from Fig.1(g) and (h),
in $\Delta m_1^2$ and $\Delta m_2^2$.
For $\Delta \lambda_5$ specifically, the  squarks  loop contributions are 
suppressed by the naturally heavier soft mass scale.
Supersymmetric models with heavy fermion loop
giving the necessary $\Delta \lambda_5$,
however, would have problem suppressing the corresponding
superparticle contributions.
While the kind of detailed analysis for the MSSM
is abundant\cite{cpc}, not much attention has been put into 
a CP violating scenario. The latter situation deserves more efforts.

We have presented a consistent 1-loop analysis of the feasibility
of radiatively induced SCPV, for both  MSSM and our proposed
new model. The latter is illustrated to have a good chance to be
phenomenologically viable. We will report more on the topic in a
forthcoming publication.

{\it Acknowledgements.} O.K. is indebted to  P.H. Frampton for discussions
and encouragement, and for suggestions on improving the manuscript. 
F.L. wants to thank L.N. Chang 
and T. Takeuchi for discussions in the early phase of this work.
O.K. was supported  by the U.S. Department of 
Energy under Grant DE-FG05-85ER-40219, Task B, and
a UNC dissertation fellowship.
F.L. was partly supported by the U.S. Department of 
Energy under Grant DE-FG05-92ER-40709, Task A.


\newpage

{\bf Figure Captions.}\\

Fig.1 : Gaugino/higgsino-loop diagrams giving rise to modifications 
to parameters in the potential. (Each dot indicates a helicity flip 
in the fermion propagator.)

Fig.2 :  Plots of radiative correction from chargino loop 
verses $m(=M_{\tilde{g}}/\mu)$, with mass mixing from EW symmetry neglected. 
(a)$\Delta m_3^2$($\overline{MS}$) in $25\times g^2\mu^2/16\pi^2$; 
(b)$\Delta \lambda_5$; (c)$\Delta \lambda_6$ ($=\Delta \lambda_7$);
both of the latter curves with values in $g^4/16\pi^2$. 

Fig.3 :  Plots of radiative correction from $t$-$T$ loop
verses $m(=M_Q/v_2)$.  (a)$\Delta m_3^2$($\overline{MS}$) 
in $10\times 3\lambda_t^2v_2^2/16\pi^2$; (b)$\Delta \lambda_5$; 
(c)$\Delta \lambda_6$; (d)$\Delta \lambda_7$; all of the latter three curves with values in $3\lambda_t^4/16\pi^2$; ($\lambda_t=\lambda_Q$, $\tan{\beta}=1$ assumed).



\begin{figure}

\includegraphics{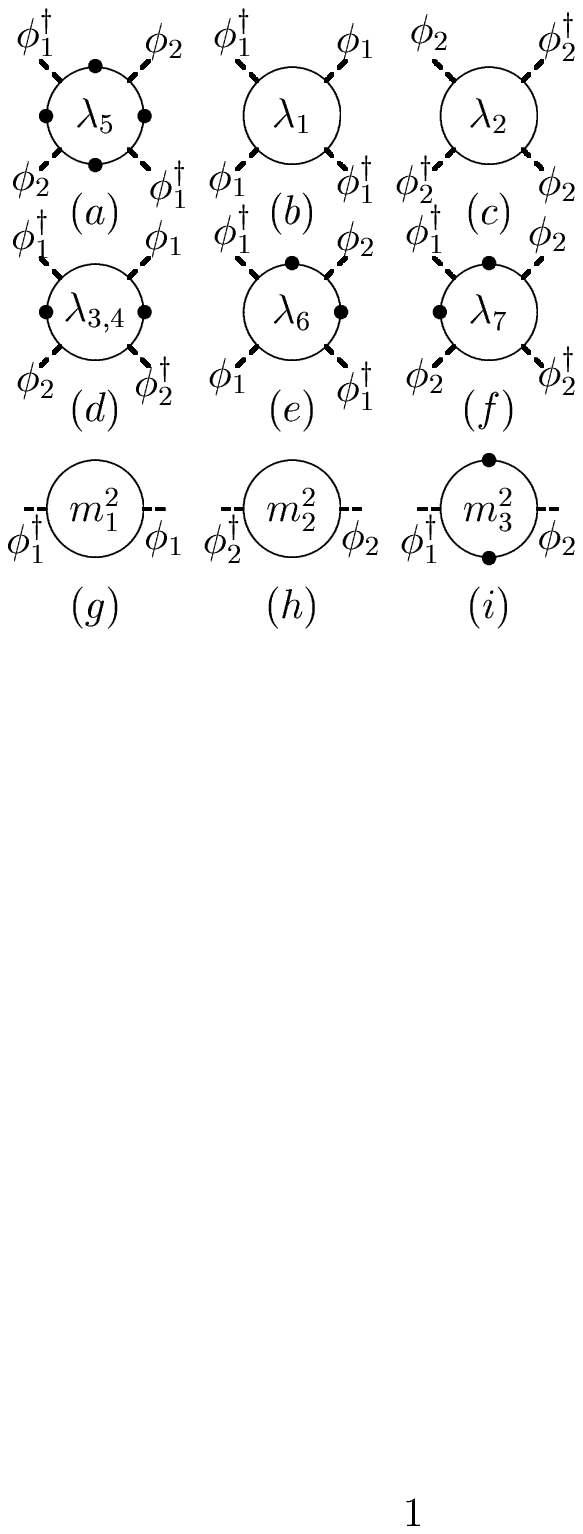}
\vspace{1in}
\caption{}
\end{figure}

\clearpage

\begin{figure}

\vspace{10in}
\includegraphics{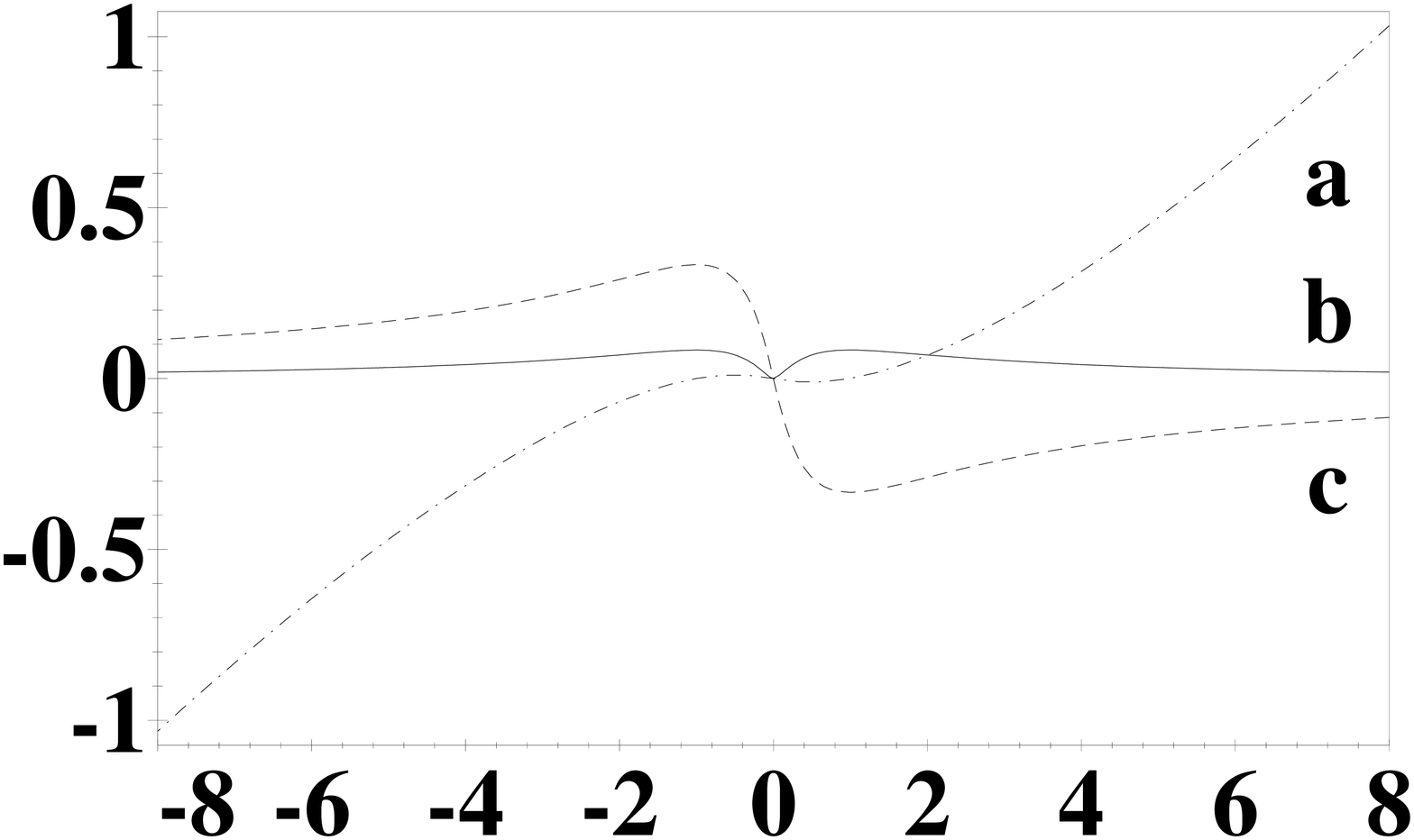}
\vspace{1in}
\caption{}
\end{figure}

\clearpage

\begin{figure}

\vspace{10in}
\includegraphics{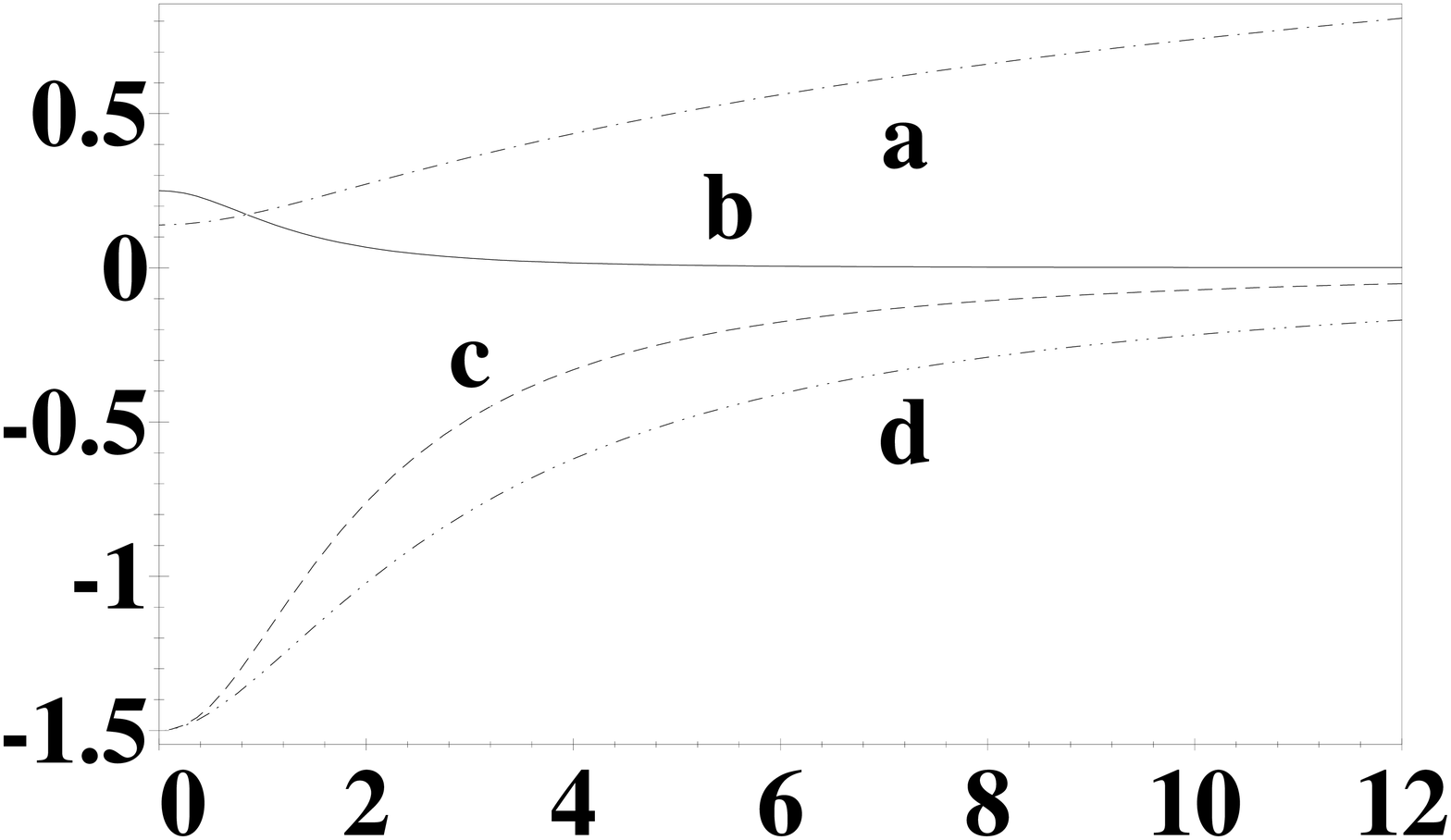}
\vspace{1in}
\caption{} 
\end{figure}

\end{document}